\documentstyle[12pt]{article}
\begin{document}

\immediate\write16{<<WARNING: LINEDRAW macros work with emTeX-dvivers
		    and other drivers supporting emTeX \special's
		    (dviscr, dvihplj, dvidot, dvips, dviwin, etc.) >>}
\newdimen\Lengthunit	   \Lengthunit	= 1.5cm
\newcount\Nhalfperiods	   \Nhalfperiods= 9
\newcount\magnitude	   \magnitude = 1000

\catcode`\*=11
\newdimen\L*   \newdimen\d*   \newdimen\d**
\newdimen\dm*  \newdimen\dd*  \newdimen\dt*
\newdimen\a*   \newdimen\b*   \newdimen\c*
\newdimen\a**  \newdimen\b**
\newdimen\xL*  \newdimen\yL*
\newdimen\rx*  \newdimen\ry*
\newdimen\tmp* \newdimen\linwid*

\newcount\k*   \newcount\l*   \newcount\m*
\newcount\k**  \newcount\l**  \newcount\m**
\newcount\n*   \newcount\dn*  \newcount\r*
\newcount\N*   \newcount\*one \newcount\*two  \*one=1 \*two=2
\newcount\*ths \*ths=1000
\newcount\angle*  \newcount\q*	\newcount\q**
\newcount\angle** \angle**=0
\newcount\sc*	  \sc*=0

\newtoks\cos*  \cos*={1}
\newtoks\sin*  \sin*={0}

\catcode`\[=13

\def\rotate(#1){\advance\angle**#1\angle*=\angle**
\q**=\angle*\ifnum\q**<0\q**=-\q**\fi
\ifnum\q**>360\q*=\angle*\divide\q*360\multiply\q*360\advance\angle*-\q*\fi
\ifnum\angle*<0\advance\angle*360\fi\q**=\angle*\divide\q**90\q**=\q**
\def\sgcos*{+}\def\sgsin*{+}\relax
\ifcase\q**\or
 \def\sgcos*{-}\def\sgsin*{+}\or
 \def\sgcos*{-}\def\sgsin*{-}\or
 \def\sgcos*{+}\def\sgsin*{-}\else\fi
\q*=\q**
\multiply\q*90\advance\angle*-\q*
\ifnum\angle*>45\sc*=1\angle*=-\angle*\advance\angle*90\else\sc*=0\fi
\def[##1,##2]{\ifnum\sc*=0\relax
\edef\cs*{\sgcos*.##1}\edef\sn*{\sgsin*.##2}\ifcase\q**\or
 \edef\cs*{\sgcos*.##2}\edef\sn*{\sgsin*.##1}\or
 \edef\cs*{\sgcos*.##1}\edef\sn*{\sgsin*.##2}\or
 \edef\cs*{\sgcos*.##2}\edef\sn*{\sgsin*.##1}\else\fi\else
\edef\cs*{\sgcos*.##2}\edef\sn*{\sgsin*.##1}\ifcase\q**\or
 \edef\cs*{\sgcos*.##1}\edef\sn*{\sgsin*.##2}\or
 \edef\cs*{\sgcos*.##2}\edef\sn*{\sgsin*.##1}\or
 \edef\cs*{\sgcos*.##1}\edef\sn*{\sgsin*.##2}\else\fi\fi
\cos*={\cs*}\sin*={\sn*}\global\edef\gcos*{\cs*}\global\edef\gsin*{\sn*}}\relax
\ifcase\angle*[9999,0]\or
[999,017]\or[999,034]\or[998,052]\or[997,069]\or[996,087]\or
[994,104]\or[992,121]\or[990,139]\or[987,156]\or[984,173]\or
[981,190]\or[978,207]\or[974,224]\or[970,241]\or[965,258]\or
[961,275]\or[956,292]\or[951,309]\or[945,325]\or[939,342]\or
[933,358]\or[927,374]\or[920,390]\or[913,406]\or[906,422]\or
[898,438]\or[891,453]\or[882,469]\or[874,484]\or[866,499]\or
[857,515]\or[848,529]\or[838,544]\or[829,559]\or[819,573]\or
[809,587]\or[798,601]\or[788,615]\or[777,629]\or[766,642]\or
[754,656]\or[743,669]\or[731,681]\or[719,694]\or[707,707]\or
\else[9999,0]\fi}

\catcode`\[=12

\def\GRAPH(hsize=#1)#2{\hbox to #1\Lengthunit{#2\hss}}

\def\Linewidth#1{\global\linwid*=#1\relax
\global\divide\linwid*10\global\multiply\linwid*\mag
\global\divide\linwid*100\special{em:linewidth \the\linwid*}}

\Linewidth{.4pt}
\def\sm*{\special{em:moveto}}
\def\sl*{\special{em:lineto}}
\let\moveto=\sm*
\let\lineto=\sl*
\newbox\spm*   \newbox\spl*
\setbox\spm*\hbox{\sm*}
\setbox\spl*\hbox{\sl*}

\def\mov#1(#2,#3)#4{\rlap{\L*=#1\Lengthunit
\xL*=#2\L* \yL*=#3\L*
\xL*=\xscale\xL* \yL*=\yscale\yL*
\rx* \the\cos*\xL* \tmp* \the\sin*\yL* \advance\rx*-\tmp*
\ry* \the\cos*\yL* \tmp* \the\sin*\xL* \advance\ry*\tmp*
\kern\rx*\raise\ry*\hbox{#4}}}

\def\rmov*(#1,#2)#3{\rlap{\xL*=#1\yL*=#2\relax
\rx* \the\cos*\xL* \tmp* \the\sin*\yL* \advance\rx*-\tmp*
\ry* \the\cos*\yL* \tmp* \the\sin*\xL* \advance\ry*\tmp*
\kern\rx*\raise\ry*\hbox{#3}}}

\def\lin#1(#2,#3){\rlap{\sm*\mov#1(#2,#3){\sl*}}}

\def\arr*(#1,#2,#3){\rmov*(#1\dd*,#1\dt*){\sm*
\rmov*(#2\dd*,#2\dt*){\rmov*(#3\dt*,-#3\dd*){\sl*}}\sm*
\rmov*(#2\dd*,#2\dt*){\rmov*(-#3\dt*,#3\dd*){\sl*}}}}

\def\arrow#1(#2,#3){\rlap{\lin#1(#2,#3)\mov#1(#2,#3){\relax
\d**=-.012\Lengthunit\dd*=#2\d**\dt*=#3\d**
\arr*(1,10,4)\arr*(3,8,4)\arr*(4.8,4.2,3)}}}

\def\arrlin#1(#2,#3){\rlap{\L*=#1\Lengthunit\L*=.5\L*
\lin#1(#2,#3)\rmov*(#2\L*,#3\L*){\arrow.1(#2,#3)}}}

\def\dasharrow#1(#2,#3){\rlap{{\Lengthunit=0.9\Lengthunit
\dashlin#1(#2,#3)\mov#1(#2,#3){\sm*}}\mov#1(#2,#3){\sl*
\d**=-.012\Lengthunit\dd*=#2\d**\dt*=#3\d**
\arr*(1,10,4)\arr*(3,8,4)\arr*(4.8,4.2,3)}}}

\def\clap#1{\hbox to 0pt{\hss #1\hss}}

\def\ind(#1,#2)#3{\rlap{\L*=.1\Lengthunit
\xL*=#1\L* \yL*=#2\L*
\rx* \the\cos*\xL* \tmp* \the\sin*\yL* \advance\rx*-\tmp*
\ry* \the\cos*\yL* \tmp* \the\sin*\xL* \advance\ry*\tmp*
\kern\rx*\raise\ry*\hbox{\lower2pt\clap{$#3$}}}}

\def\sh*(#1,#2)#3{\rlap{\dm*=\the\n*\d**
\xL*=\xscale\dm* \yL*=\yscale\dm* \xL*=#1\xL* \yL*=#2\yL*
\rx* \the\cos*\xL* \tmp* \the\sin*\yL* \advance\rx*-\tmp*
\ry* \the\cos*\yL* \tmp* \the\sin*\xL* \advance\ry*\tmp*
\kern\rx*\raise\ry*\hbox{#3}}}

\def\calcnum*#1(#2,#3){\a*=1000sp\b*=1000sp\a*=#2\a*\b*=#3\b*
\ifdim\a*<0pt\a*-\a*\fi\ifdim\b*<0pt\b*-\b*\fi
\ifdim\a*>\b*\c*=.96\a*\advance\c*.4\b*
\else\c*=.96\b*\advance\c*.4\a*\fi
\k*\a*\multiply\k*\k*\l*\b*\multiply\l*\l*
\m*\k*\advance\m*\l*\n*\c*\r*\n*\multiply\n*\n*
\dn*\m*\advance\dn*-\n*\divide\dn*2\divide\dn*\r*
\advance\r*\dn*
\c*=\the\Nhalfperiods5sp\c*=#1\c*\ifdim\c*<0pt\c*-\c*\fi
\multiply\c*\r*\N*\c*\divide\N*10000}

\def\dashlin#1(#2,#3){\rlap{\calcnum*#1(#2,#3)\relax
\d**=#1\Lengthunit\ifdim\d**<0pt\d**-\d**\fi
\divide\N*2\multiply\N*2\advance\N*\*one
\divide\d**\N*\sm*\n*\*one\sh*(#2,#3){\sl*}\loop
\advance\n*\*one\sh*(#2,#3){\sm*}\advance\n*\*one
\sh*(#2,#3){\sl*}\ifnum\n*<\N*\repeat}}

\def\dashdotlin#1(#2,#3){\rlap{\calcnum*#1(#2,#3)\relax
\d**=#1\Lengthunit\ifdim\d**<0pt\d**-\d**\fi
\divide\N*2\multiply\N*2\advance\N*1\multiply\N*2\relax
\divide\d**\N*\sm*\n*\*two\sh*(#2,#3){\sl*}\loop
\advance\n*\*one\sh*(#2,#3){\kern-1.48pt\lower.5pt\hbox{\rm.}}\relax
\advance\n*\*one\sh*(#2,#3){\sm*}\advance\n*\*two
\sh*(#2,#3){\sl*}\ifnum\n*<\N*\repeat}}

\def\shl*(#1,#2)#3{\kern#1#3\lower#2#3\hbox{\unhcopy\spl*}}

\def\trianglin#1(#2,#3){\rlap{\toks0={#2}\toks1={#3}\calcnum*#1(#2,#3)\relax
\dd*=.57\Lengthunit\dd*=#1\dd*\divide\dd*\N*
\divide\dd*\*ths \multiply\dd*\magnitude
\d**=#1\Lengthunit\ifdim\d**<0pt\d**-\d**\fi
\multiply\N*2\divide\d**\N*\sm*\n*\*one\loop
\shl**{\dd*}\dd*-\dd*\advance\n*2\relax
\ifnum\n*<\N*\repeat\n*\N*\shl**{0pt}}}

\def\wavelin#1(#2,#3){\rlap{\toks0={#2}\toks1={#3}\calcnum*#1(#2,#3)\relax
\dd*=.23\Lengthunit\dd*=#1\dd*\divide\dd*\N*
\divide\dd*\*ths \multiply\dd*\magnitude
\d**=#1\Lengthunit\ifdim\d**<0pt\d**-\d**\fi
\multiply\N*4\divide\d**\N*\sm*\n*\*one\loop
\shl**{\dd*}\dt*=1.3\dd*\advance\n*\*one
\shl**{\dt*}\advance\n*\*one
\shl**{\dd*}\advance\n*\*two
\dd*-\dd*\ifnum\n*<\N*\repeat\n*\N*\shl**{0pt}}}

\def\w*lin(#1,#2){\rlap{\toks0={#1}\toks1={#2}\d**=\Lengthunit\dd*=-.12\d**
\divide\dd*\*ths \multiply\dd*\magnitude
\N*8\divide\d**\N*\sm*\n*\*one\loop
\shl**{\dd*}\dt*=1.3\dd*\advance\n*\*one
\shl**{\dt*}\advance\n*\*one
\shl**{\dd*}\advance\n*\*one
\shl**{0pt}\dd*-\dd*\advance\n*1\ifnum\n*<\N*\repeat}}

\def\l*arc(#1,#2)[#3][#4]{\rlap{\toks0={#1}\toks1={#2}\d**=\Lengthunit
\dd*=#3.037\d**\dd*=#4\dd*\dt*=#3.049\d**\dt*=#4\dt*\ifdim\d**>10mm\relax
\d**=.25\d**\n*\*one\shl**{-\dd*}\n*\*two\shl**{-\dt*}\n*3\relax
\shl**{-\dd*}\n*4\relax\shl**{0pt}\else
\ifdim\d**>5mm\d**=.5\d**\n*\*one\shl**{-\dt*}\n*\*two
\shl**{0pt}\else\n*\*one\shl**{0pt}\fi\fi}}

\def\d*arc(#1,#2)[#3][#4]{\rlap{\toks0={#1}\toks1={#2}\d**=\Lengthunit
\dd*=#3.037\d**\dd*=#4\dd*\d**=.25\d**\sm*\n*\*one\shl**{-\dd*}\relax
\n*3\relax\sh*(#1,#2){\xL*=\xscale\dd*\yL*=\yscale\dd*
\kern#2\xL*\lower#1\yL*\hbox{\sm*}}\n*4\relax\shl**{0pt}}}

\def\shl**#1{\c*=\the\n*\d**\d*=#1\relax
\a*=\the\toks0\c*\b*=\the\toks1\d*\advance\a*-\b*
\b*=\the\toks1\c*\d*=\the\toks0\d*\advance\b*\d*
\a*=\xscale\a*\b*=\yscale\b*
\rx* \the\cos*\a* \tmp* \the\sin*\b* \advance\rx*-\tmp*
\ry* \the\cos*\b* \tmp* \the\sin*\a* \advance\ry*\tmp*
\raise\ry*\rlap{\kern\rx*\unhcopy\spl*}}

\def\wlin*#1(#2,#3)[#4]{\rlap{\toks0={#2}\toks1={#3}\relax
\c*=#1\l*\c*\c*=.01\Lengthunit\m*\c*\divide\l*\m*
\c*=\the\Nhalfperiods5sp\multiply\c*\l*\N*\c*\divide\N*\*ths
\divide\N*2\multiply\N*2\advance\N*\*one
\dd*=.002\Lengthunit\dd*=#4\dd*\multiply\dd*\l*\divide\dd*\N*
\divide\dd*\*ths \multiply\dd*\magnitude
\d**=#1\multiply\N*4\divide\d**\N*\sm*\n*\*one\loop
\shl**{\dd*}\dt*=1.3\dd*\advance\n*\*one
\shl**{\dt*}\advance\n*\*one
\shl**{\dd*}\advance\n*\*two
\dd*-\dd*\ifnum\n*<\N*\repeat\n*\N*\shl**{0pt}}}

\def\wavebox#1{\setbox0\hbox{#1}\relax
\a*=\wd0\advance\a*14pt\b*=\ht0\advance\b*\dp0\advance\b*14pt\relax
\hbox{\kern9pt\relax
\rmov*(0pt,\ht0){\rmov*(-7pt,7pt){\wlin*\a*(1,0)[+]\wlin*\b*(0,-1)[-]}}\relax
\rmov*(\wd0,-\dp0){\rmov*(7pt,-7pt){\wlin*\a*(-1,0)[+]\wlin*\b*(0,1)[-]}}\relax
\box0\kern9pt}}

\def\rectangle#1(#2,#3){\relax
\lin#1(#2,0)\lin#1(0,#3)\mov#1(0,#3){\lin#1(#2,0)}\mov#1(#2,0){\lin#1(0,#3)}}

\def\dashrectangle#1(#2,#3){\dashlin#1(#2,0)\dashlin#1(0,#3)\relax
\mov#1(0,#3){\dashlin#1(#2,0)}\mov#1(#2,0){\dashlin#1(0,#3)}}

\def\waverectangle#1(#2,#3){\L*=#1\Lengthunit\a*=#2\L*\b*=#3\L*
\ifdim\a*<0pt\a*-\a*\def\x*{-1}\else\def\x*{1}\fi
\ifdim\b*<0pt\b*-\b*\def\y*{-1}\else\def\y*{1}\fi
\wlin*\a*(\x*,0)[-]\wlin*\b*(0,\y*)[+]\relax
\mov#1(0,#3){\wlin*\a*(\x*,0)[+]}\mov#1(#2,0){\wlin*\b*(0,\y*)[-]}}

\def\calcparab*{\ifnum\n*>\m*\k*\N*\advance\k*-\n*\else\k*\n*\fi
\a*=\the\k* sp\a*=10\a*\b*\dm*\advance\b*-\a*\k*\b*
\a*=\the\*ths\b*\divide\a*\l*\multiply\a*\k*
\divide\a*\l*\k*\*ths\r*\a*\advance\k*-\r*\dt*=\the\k*\L*}

\def\arcto#1(#2,#3)[#4]{\rlap{\toks0={#2}\toks1={#3}\calcnum*#1(#2,#3)\relax
\dm*=135sp\dm*=#1\dm*\d**=#1\Lengthunit\ifdim\dm*<0pt\dm*-\dm*\fi
\multiply\dm*\r*\a*=.3\dm*\a*=#4\a*\ifdim\a*<0pt\a*-\a*\fi
\advance\dm*\a*\N*\dm*\divide\N*10000\relax
\divide\N*2\multiply\N*2\advance\N*\*one
\L*=-.25\d**\L*=#4\L*\divide\d**\N*\divide\L*\*ths
\m*\N*\divide\m*2\dm*=\the\m*5sp\l*\dm*\sm*\n*\*one\loop
\calcparab*\shl**{-\dt*}\advance\n*1\ifnum\n*<\N*\repeat}}

\def\arrarcto#1(#2,#3)[#4]{\L*=#1\Lengthunit\L*=.54\L*
\arcto#1(#2,#3)[#4]\rmov*(#2\L*,#3\L*){\d*=.457\L*\d*=#4\d*\d**-\d*
\rmov*(#3\d**,#2\d*){\arrow.02(#2,#3)}}}

\def\dasharcto#1(#2,#3)[#4]{\rlap{\toks0={#2}\toks1={#3}\relax
\calcnum*#1(#2,#3)\dm*=\the\N*5sp\a*=.3\dm*\a*=#4\a*\ifdim\a*<0pt\a*-\a*\fi
\advance\dm*\a*\N*\dm*
\divide\N*20\multiply\N*2\advance\N*1\d**=#1\Lengthunit
\L*=-.25\d**\L*=#4\L*\divide\d**\N*\divide\L*\*ths
\m*\N*\divide\m*2\dm*=\the\m*5sp\l*\dm*
\sm*\n*\*one\loop\calcparab*
\shl**{-\dt*}\advance\n*1\ifnum\n*>\N*\else\calcparab*
\sh*(#2,#3){\xL*=#3\dt* \yL*=#2\dt*
\rx* \the\cos*\xL* \tmp* \the\sin*\yL* \advance\rx*\tmp*
\ry* \the\cos*\yL* \tmp* \the\sin*\xL* \advance\ry*-\tmp*
\kern\rx*\lower\ry*\hbox{\sm*}}\fi
\advance\n*1\ifnum\n*<\N*\repeat}}

\def\*shl*#1{\c*=\the\n*\d**\advance\c*#1\a**\d*\dt*\advance\d*#1\b**
\a*=\the\toks0\c*\b*=\the\toks1\d*\advance\a*-\b*
\b*=\the\toks1\c*\d*=\the\toks0\d*\advance\b*\d*
\rx* \the\cos*\a* \tmp* \the\sin*\b* \advance\rx*-\tmp*
\ry* \the\cos*\b* \tmp* \the\sin*\a* \advance\ry*\tmp*
\raise\ry*\rlap{\kern\rx*\unhcopy\spl*}}

\def\calcnormal*#1{\b**=10000sp\a**\b**\k*\n*\advance\k*-\m*
\multiply\a**\k*\divide\a**\m*\a**=#1\a**\ifdim\a**<0pt\a**-\a**\fi
\ifdim\a**>\b**\d*=.96\a**\advance\d*.4\b**
\else\d*=.96\b**\advance\d*.4\a**\fi
\d*=.01\d*\r*\d*\divide\a**\r*\divide\b**\r*
\ifnum\k*<0\a**-\a**\fi\d*=#1\d*\ifdim\d*<0pt\b**-\b**\fi
\k*\a**\a**=\the\k*\dd*\k*\b**\b**=\the\k*\dd*}

\def\wavearcto#1(#2,#3)[#4]{\rlap{\toks0={#2}\toks1={#3}\relax
\calcnum*#1(#2,#3)\c*=\the\N*5sp\a*=.4\c*\a*=#4\a*\ifdim\a*<0pt\a*-\a*\fi
\advance\c*\a*\N*\c*\divide\N*20\multiply\N*2\advance\N*-1\multiply\N*4\relax
\d**=#1\Lengthunit\dd*=.012\d**
\divide\dd*\*ths \multiply\dd*\magnitude
\ifdim\d**<0pt\d**-\d**\fi\L*=.25\d**
\divide\d**\N*\divide\dd*\N*\L*=#4\L*\divide\L*\*ths
\m*\N*\divide\m*2\dm*=\the\m*0sp\l*\dm*
\sm*\n*\*one\loop\calcnormal*{#4}\calcparab*
\*shl*{1}\advance\n*\*one\calcparab*
\*shl*{1.3}\advance\n*\*one\calcparab*
\*shl*{1}\advance\n*2\dd*-\dd*\ifnum\n*<\N*\repeat\n*\N*\shl**{0pt}}}

\def\triangarcto#1(#2,#3)[#4]{\rlap{\toks0={#2}\toks1={#3}\relax
\calcnum*#1(#2,#3)\c*=\the\N*5sp\a*=.4\c*\a*=#4\a*\ifdim\a*<0pt\a*-\a*\fi
\advance\c*\a*\N*\c*\divide\N*20\multiply\N*2\advance\N*-1\multiply\N*2\relax
\d**=#1\Lengthunit\dd*=.012\d**
\divide\dd*\*ths \multiply\dd*\magnitude
\ifdim\d**<0pt\d**-\d**\fi\L*=.25\d**
\divide\d**\N*\divide\dd*\N*\L*=#4\L*\divide\L*\*ths
\m*\N*\divide\m*2\dm*=\the\m*0sp\l*\dm*
\sm*\n*\*one\loop\calcnormal*{#4}\calcparab*
\*shl*{1}\advance\n*2\dd*-\dd*\ifnum\n*<\N*\repeat\n*\N*\shl**{0pt}}}

\def\hr*#1{\L*=\xscale\Lengthunit\ifnum
\angle**=0\clap{\vrule width#1\L* height.1pt}\else
\L*=#1\L*\L*=.5\L*\rmov*(-\L*,0pt){\sm*}\rmov*(\L*,0pt){\sl*}\fi}

\def\shade#1[#2]{\rlap{\Lengthunit=#1\Lengthunit
\special{em:linewidth .001pt}\relax
\mov(0,#2.05){\hr*{.994}}\mov(0,#2.1){\hr*{.980}}\relax
\mov(0,#2.15){\hr*{.953}}\mov(0,#2.2){\hr*{.916}}\relax
\mov(0,#2.25){\hr*{.867}}\mov(0,#2.3){\hr*{.798}}\relax
\mov(0,#2.35){\hr*{.715}}\mov(0,#2.4){\hr*{.603}}\relax
\mov(0,#2.45){\hr*{.435}}\special{em:linewidth \the\linwid*}}}

\def\dshade#1[#2]{\rlap{\special{em:linewidth .001pt}\relax
\Lengthunit=#1\Lengthunit\if#2-\def\t*{+}\else\def\t*{-}\fi
\mov(0,\t*.025){\relax
\mov(0,#2.05){\hr*{.995}}\mov(0,#2.1){\hr*{.988}}\relax
\mov(0,#2.15){\hr*{.969}}\mov(0,#2.2){\hr*{.937}}\relax
\mov(0,#2.25){\hr*{.893}}\mov(0,#2.3){\hr*{.836}}\relax
\mov(0,#2.35){\hr*{.760}}\mov(0,#2.4){\hr*{.662}}\relax
\mov(0,#2.45){\hr*{.531}}\mov(0,#2.5){\hr*{.320}}\relax
\special{em:linewidth \the\linwid*}}}}

\def\vdot{\rlap{\kern-1.9pt\lower1.8pt\hbox{$\scriptstyle\bullet$}}}
\def\vtimes{\rlap{\kern-3pt\lower1.8pt\hbox{$\scriptstyle\times$}}}
\def\vDot{\rlap{\kern-2.3pt\lower2.7pt\hbox{$\bullet$}}}
\def\vTimes{\rlap{\kern-3.6pt\lower2.4pt\hbox{$\times$}}}

\def\arc(#1)[#2,#3]{{\k*=#2\l*=#3\m*=\l*
\advance\m*-6\ifnum\k*>\l*\relax\else
{\rotate(#2)\mov(#1,0){\sm*}}\loop
\ifnum\k*<\m*\advance\k*5{\rotate(\k*)\mov(#1,0){\sl*}}\repeat
{\rotate(#3)\mov(#1,0){\sl*}}\fi}}

\def\dasharc(#1)[#2,#3]{{\k**=#2\n*=#3\advance\n*-1\advance\n*-\k**
\L*=1000sp\L*#1\L* \multiply\L*\n* \multiply\L*\Nhalfperiods
\divide\L*57\N*\L* \divide\N*2000\ifnum\N*=0\N*1\fi
\r*\n*	\divide\r*\N* \ifnum\r*<2\r*2\fi
\m**\r* \divide\m**2 \l**\r* \advance\l**-\m** \N*\n* \divide\N*\r*
\k**\r* \multiply\k**\N* \dn*\n* \advance\dn*-\k** \divide\dn*2\advance\dn*\*one
\r*\l** \divide\r*2\advance\dn*\r* \advance\N*-2\k**#2\relax
\ifnum\l**<6{\rotate(#2)\mov(#1,0){\sm*}}\advance\k**\dn*
{\rotate(\k**)\mov(#1,0){\sl*}}\advance\k**\m**
{\rotate(\k**)\mov(#1,0){\sm*}}\loop
\advance\k**\l**{\rotate(\k**)\mov(#1,0){\sl*}}\advance\k**\m**
{\rotate(\k**)\mov(#1,0){\sm*}}\advance\N*-1\ifnum\N*>0\repeat
{\rotate(#3)\mov(#1,0){\sl*}}\else\advance\k**\dn*
\arc(#1)[#2,\k**]\loop\advance\k**\m** \r*\k**
\advance\k**\l** {\arc(#1)[\r*,\k**]}\relax
\advance\N*-1\ifnum\N*>0\repeat
\advance\k**\m**\arc(#1)[\k**,#3]\fi}}

\def\triangarc#1(#2)[#3,#4]{{\k**=#3\n*=#4\advance\n*-\k**
\L*=1000sp\L*#2\L* \multiply\L*\n* \multiply\L*\Nhalfperiods
\divide\L*57\N*\L* \divide\N*1000\ifnum\N*=0\N*1\fi
\d**=#2\Lengthunit \d*\d** \divide\d*57\multiply\d*\n*
\r*\n*	\divide\r*\N* \ifnum\r*<2\r*2\fi
\m**\r* \divide\m**2 \l**\r* \advance\l**-\m** \N*\n* \divide\N*\r*
\dt*\d* \divide\dt*\N* \dt*.5\dt* \dt*#1\dt*
\divide\dt*1000\multiply\dt*\magnitude
\k**\r* \multiply\k**\N* \dn*\n* \advance\dn*-\k** \divide\dn*2\relax
\r*\l** \divide\r*2\advance\dn*\r* \advance\N*-1\k**#3\relax
{\rotate(#3)\mov(#2,0){\sm*}}\advance\k**\dn*
{\rotate(\k**)\mov(#2,0){\sl*}}\advance\k**-\m**\advance\l**\m**\loop\dt*-\dt*
\d*\d** \advance\d*\dt*
\advance\k**\l**{\rotate(\k**)\rmov*(\d*,0pt){\sl*}}%
\advance\N*-1\ifnum\N*>0\repeat\advance\k**\m**
{\rotate(\k**)\mov(#2,0){\sl*}}{\rotate(#4)\mov(#2,0){\sl*}}}}

\def\wavearc#1(#2)[#3,#4]{{\k**=#3\n*=#4\advance\n*-\k**
\L*=4000sp\L*#2\L* \multiply\L*\n* \multiply\L*\Nhalfperiods
\divide\L*57\N*\L* \divide\N*1000\ifnum\N*=0\N*1\fi
\d**=#2\Lengthunit \d*\d** \divide\d*57\multiply\d*\n*
\r*\n*	\divide\r*\N* \ifnum\r*=0\r*1\fi
\m**\r* \divide\m**2 \l**\r* \advance\l**-\m** \N*\n* \divide\N*\r*
\dt*\d* \divide\dt*\N* \dt*.7\dt* \dt*#1\dt*
\divide\dt*1000\multiply\dt*\magnitude
\k**\r* \multiply\k**\N* \dn*\n* \advance\dn*-\k** \divide\dn*2\relax
\divide\N*4\advance\N*-1\k**#3\relax
{\rotate(#3)\mov(#2,0){\sm*}}\advance\k**\dn*
{\rotate(\k**)\mov(#2,0){\sl*}}\advance\k**-\m**\advance\l**\m**\loop\dt*-\dt*
\d*\d** \advance\d*\dt* \dd*\d** \advance\dd*1.3\dt*
\advance\k**\r*{\rotate(\k**)\rmov*(\d*,0pt){\sl*}}\relax
\advance\k**\r*{\rotate(\k**)\rmov*(\dd*,0pt){\sl*}}\relax
\advance\k**\r*{\rotate(\k**)\rmov*(\d*,0pt){\sl*}}\relax
\advance\k**\r*
\advance\N*-1\ifnum\N*>0\repeat\advance\k**\m**
{\rotate(\k**)\mov(#2,0){\sl*}}{\rotate(#4)\mov(#2,0){\sl*}}}}

\def\gmov*#1(#2,#3)#4{\rlap{\L*=#1\Lengthunit
\xL*=#2\L* \yL*=#3\L*
\rx* \gcos*\xL* \tmp* \gsin*\yL* \advance\rx*-\tmp*
\ry* \gcos*\yL* \tmp* \gsin*\xL* \advance\ry*\tmp*
\rx*=\xscale\rx* \ry*=\yscale\ry*
\xL* \the\cos*\rx* \tmp* \the\sin*\ry* \advance\xL*-\tmp*
\yL* \the\cos*\ry* \tmp* \the\sin*\rx* \advance\yL*\tmp*
\kern\xL*\raise\yL*\hbox{#4}}}

\def\rgmov*(#1,#2)#3{\rlap{\xL*#1\yL*#2\relax
\rx* \gcos*\xL* \tmp* \gsin*\yL* \advance\rx*-\tmp*
\ry* \gcos*\yL* \tmp* \gsin*\xL* \advance\ry*\tmp*
\rx*=\xscale\rx* \ry*=\yscale\ry*
\xL* \the\cos*\rx* \tmp* \the\sin*\ry* \advance\xL*-\tmp*
\yL* \the\cos*\ry* \tmp* \the\sin*\rx* \advance\yL*\tmp*
\kern\xL*\raise\yL*\hbox{#3}}}

\def\Earc(#1)[#2,#3][#4,#5]{{\k*=#2\l*=#3\m*=\l*
\advance\m*-6\ifnum\k*>\l*\relax\else\def\xscale{#4}\def\yscale{#5}\relax
{\angle**0\rotate(#2)}\gmov*(#1,0){\sm*}\loop
\ifnum\k*<\m*\advance\k*5\relax
{\angle**0\rotate(\k*)}\gmov*(#1,0){\sl*}\repeat
{\angle**0\rotate(#3)}\gmov*(#1,0){\sl*}\relax
\def\xscale{1}\def\yscale{1}\fi}}

\def\dashEarc(#1)[#2,#3][#4,#5]{{\k**=#2\n*=#3\advance\n*-1\advance\n*-\k**
\L*=1000sp\L*#1\L* \multiply\L*\n* \multiply\L*\Nhalfperiods
\divide\L*57\N*\L* \divide\N*2000\ifnum\N*=0\N*1\fi
\r*\n*	\divide\r*\N* \ifnum\r*<2\r*2\fi
\m**\r* \divide\m**2 \l**\r* \advance\l**-\m** \N*\n* \divide\N*\r*
\k**\r*\multiply\k**\N* \dn*\n* \advance\dn*-\k** \divide\dn*2\advance\dn*\*one
\r*\l** \divide\r*2\advance\dn*\r* \advance\N*-2\k**#2\relax
\ifnum\l**<6\def\xscale{#4}\def\yscale{#5}\relax
{\angle**0\rotate(#2)}\gmov*(#1,0){\sm*}\advance\k**\dn*
{\angle**0\rotate(\k**)}\gmov*(#1,0){\sl*}\advance\k**\m**
{\angle**0\rotate(\k**)}\gmov*(#1,0){\sm*}\loop
\advance\k**\l**{\angle**0\rotate(\k**)}\gmov*(#1,0){\sl*}\advance\k**\m**
{\angle**0\rotate(\k**)}\gmov*(#1,0){\sm*}\advance\N*-1\ifnum\N*>0\repeat
{\angle**0\rotate(#3)}\gmov*(#1,0){\sl*}\def\xscale{1}\def\yscale{1}\else
\advance\k**\dn* \Earc(#1)[#2,\k**][#4,#5]\loop\advance\k**\m** \r*\k**
\advance\k**\l** {\Earc(#1)[\r*,\k**][#4,#5]}\relax
\advance\N*-1\ifnum\N*>0\repeat
\advance\k**\m**\Earc(#1)[\k**,#3][#4,#5]\fi}}

\def\triangEarc#1(#2)[#3,#4][#5,#6]{{\k**=#3\n*=#4\advance\n*-\k**
\L*=1000sp\L*#2\L* \multiply\L*\n* \multiply\L*\Nhalfperiods
\divide\L*57\N*\L* \divide\N*1000\ifnum\N*=0\N*1\fi
\d**=#2\Lengthunit \d*\d** \divide\d*57\multiply\d*\n*
\r*\n*	\divide\r*\N* \ifnum\r*<2\r*2\fi
\m**\r* \divide\m**2 \l**\r* \advance\l**-\m** \N*\n* \divide\N*\r*
\dt*\d* \divide\dt*\N* \dt*.5\dt* \dt*#1\dt*
\divide\dt*1000\multiply\dt*\magnitude
\k**\r* \multiply\k**\N* \dn*\n* \advance\dn*-\k** \divide\dn*2\relax
\r*\l** \divide\r*2\advance\dn*\r* \advance\N*-1\k**#3\relax
\def\xscale{#5}\def\yscale{#6}\relax
{\angle**0\rotate(#3)}\gmov*(#2,0){\sm*}\advance\k**\dn*
{\angle**0\rotate(\k**)}\gmov*(#2,0){\sl*}\advance\k**-\m**
\advance\l**\m**\loop\dt*-\dt* \d*\d** \advance\d*\dt*
\advance\k**\l**{\angle**0\rotate(\k**)}\rgmov*(\d*,0pt){\sl*}\relax
\advance\N*-1\ifnum\N*>0\repeat\advance\k**\m**
{\angle**0\rotate(\k**)}\gmov*(#2,0){\sl*}\relax
{\angle**0\rotate(#4)}\gmov*(#2,0){\sl*}\def\xscale{1}\def\yscale{1}}}

\def\waveEarc#1(#2)[#3,#4][#5,#6]{{\k**=#3\n*=#4\advance\n*-\k**
\L*=4000sp\L*#2\L* \multiply\L*\n* \multiply\L*\Nhalfperiods
\divide\L*57\N*\L* \divide\N*1000\ifnum\N*=0\N*1\fi
\d**=#2\Lengthunit \d*\d** \divide\d*57\multiply\d*\n*
\r*\n*	\divide\r*\N* \ifnum\r*=0\r*1\fi
\m**\r* \divide\m**2 \l**\r* \advance\l**-\m** \N*\n* \divide\N*\r*
\dt*\d* \divide\dt*\N* \dt*.7\dt* \dt*#1\dt*
\divide\dt*1000\multiply\dt*\magnitude
\k**\r* \multiply\k**\N* \dn*\n* \advance\dn*-\k** \divide\dn*2\relax
\divide\N*4\advance\N*-1\k**#3\def\xscale{#5}\def\yscale{#6}\relax
{\angle**0\rotate(#3)}\gmov*(#2,0){\sm*}\advance\k**\dn*
{\angle**0\rotate(\k**)}\gmov*(#2,0){\sl*}\advance\k**-\m**
\advance\l**\m**\loop\dt*-\dt*
\d*\d** \advance\d*\dt* \dd*\d** \advance\dd*1.3\dt*
\advance\k**\r*{\angle**0\rotate(\k**)}\rgmov*(\d*,0pt){\sl*}\relax
\advance\k**\r*{\angle**0\rotate(\k**)}\rgmov*(\dd*,0pt){\sl*}\relax
\advance\k**\r*{\angle**0\rotate(\k**)}\rgmov*(\d*,0pt){\sl*}\relax
\advance\k**\r*
\advance\N*-1\ifnum\N*>0\repeat\advance\k**\m**
{\angle**0\rotate(\k**)}\gmov*(#2,0){\sl*}\relax
{\angle**0\rotate(#4)}\gmov*(#2,0){\sl*}\def\xscale{1}\def\yscale{1}}}

\newcount\CatcodeOfAtSign
\CatcodeOfAtSign=\the\catcode`\@
\catcode`\@=11
\def\@arc#1[#2][#3]{\rlap{\Lengthunit=#1\Lengthunit
\sm*\l*arc(#2.1914,#3.0381)[#2][#3]\relax
\mov(#2.1914,#3.0381){\l*arc(#2.1622,#3.1084)[#2][#3]}\relax
\mov(#2.3536,#3.1465){\l*arc(#2.1084,#3.1622)[#2][#3]}\relax
\mov(#2.4619,#3.3086){\l*arc(#2.0381,#3.1914)[#2][#3]}}}

\def\dash@arc#1[#2][#3]{\rlap{\Lengthunit=#1\Lengthunit
\d*arc(#2.1914,#3.0381)[#2][#3]\relax
\mov(#2.1914,#3.0381){\d*arc(#2.1622,#3.1084)[#2][#3]}\relax
\mov(#2.3536,#3.1465){\d*arc(#2.1084,#3.1622)[#2][#3]}\relax
\mov(#2.4619,#3.3086){\d*arc(#2.0381,#3.1914)[#2][#3]}}}

\def\wave@arc#1[#2][#3]{\rlap{\Lengthunit=#1\Lengthunit
\w*lin(#2.1914,#3.0381)\relax
\mov(#2.1914,#3.0381){\w*lin(#2.1622,#3.1084)}\relax
\mov(#2.3536,#3.1465){\w*lin(#2.1084,#3.1622)}\relax
\mov(#2.4619,#3.3086){\w*lin(#2.0381,#3.1914)}}}

\def\bezier#1(#2,#3)(#4,#5)(#6,#7){\N*#1\l*\N* \advance\l*\*one
\d* #4\Lengthunit \advance\d* -#2\Lengthunit \multiply\d* \*two
\b* #6\Lengthunit \advance\b* -#2\Lengthunit
\advance\b*-\d* \divide\b*\N*
\d** #5\Lengthunit \advance\d** -#3\Lengthunit \multiply\d** \*two
\b** #7\Lengthunit \advance\b** -#3\Lengthunit
\advance\b** -\d** \divide\b**\N*
\mov(#2,#3){\sm*{\loop\ifnum\m*<\l*
\a*\m*\b* \advance\a*\d* \divide\a*\N* \multiply\a*\m*
\a**\m*\b** \advance\a**\d** \divide\a**\N* \multiply\a**\m*
\rmov*(\a*,\a**){\unhcopy\spl*}\advance\m*\*one\repeat}}}

\catcode`\*=12

\newcount\n@ast
\def\n@ast@#1{\n@ast0\relax\get@ast@#1\end}
\def\get@ast@#1{\ifx#1\end\let\next\relax\else
\ifx#1*\advance\n@ast1\fi\let\next\get@ast@\fi\next}

\newif\if@up \newif\if@dwn
\def\up@down@#1{\@upfalse\@dwnfalse
\if#1u\@uptrue\fi\if#1U\@uptrue\fi\if#1+\@uptrue\fi
\if#1d\@dwntrue\fi\if#1D\@dwntrue\fi\if#1-\@dwntrue\fi}

\def\halfcirc#1(#2)[#3]{{\Lengthunit=#2\Lengthunit\up@down@{#3}\relax
\if@up\mov(0,.5){\@arc[-][-]\@arc[+][-]}\fi
\if@dwn\mov(0,-.5){\@arc[-][+]\@arc[+][+]}\fi
\def\lft{\mov(0,.5){\@arc[-][-]}\mov(0,-.5){\@arc[-][+]}}\relax
\def\rght{\mov(0,.5){\@arc[+][-]}\mov(0,-.5){\@arc[+][+]}}\relax
\if#3l\lft\fi\if#3L\lft\fi\if#3r\rght\fi\if#3R\rght\fi
\n@ast@{#1}\relax
\ifnum\n@ast>0\if@up\shade[+]\fi\if@dwn\shade[-]\fi\fi
\ifnum\n@ast>1\if@up\dshade[+]\fi\if@dwn\dshade[-]\fi\fi}}

\def\halfdashcirc(#1)[#2]{{\Lengthunit=#1\Lengthunit\up@down@{#2}\relax
\if@up\mov(0,.5){\dash@arc[-][-]\dash@arc[+][-]}\fi
\if@dwn\mov(0,-.5){\dash@arc[-][+]\dash@arc[+][+]}\fi
\def\lft{\mov(0,.5){\dash@arc[-][-]}\mov(0,-.5){\dash@arc[-][+]}}\relax
\def\rght{\mov(0,.5){\dash@arc[+][-]}\mov(0,-.5){\dash@arc[+][+]}}\relax
\if#2l\lft\fi\if#2L\lft\fi\if#2r\rght\fi\if#2R\rght\fi}}

\def\halfwavecirc(#1)[#2]{{\Lengthunit=#1\Lengthunit\up@down@{#2}\relax
\if@up\mov(0,.5){\wave@arc[-][-]\wave@arc[+][-]}\fi
\if@dwn\mov(0,-.5){\wave@arc[-][+]\wave@arc[+][+]}\fi
\def\lft{\mov(0,.5){\wave@arc[-][-]}\mov(0,-.5){\wave@arc[-][+]}}\relax
\def\rght{\mov(0,.5){\wave@arc[+][-]}\mov(0,-.5){\wave@arc[+][+]}}\relax
\if#2l\lft\fi\if#2L\lft\fi\if#2r\rght\fi\if#2R\rght\fi}}

\catcode`\*=11

\def\Circle#1(#2){\halfcirc#1(#2)[u]\halfcirc#1(#2)[d]\n@ast@{#1}\relax
\ifnum\n@ast>0\L*=\xscale\Lengthunit
\ifnum\angle**=0\clap{\vrule width#2\L* height.1pt}\else
\L*=#2\L*\L*=.5\L*\special{em:linewidth .001pt}\relax
\rmov*(-\L*,0pt){\sm*}\rmov*(\L*,0pt){\sl*}\relax
\special{em:linewidth \the\linwid*}\fi\fi}

\catcode`\*=12

\def\wavecirc(#1){\halfwavecirc(#1)[u]\halfwavecirc(#1)[d]}

\def\dashcirc(#1){\halfdashcirc(#1)[u]\halfdashcirc(#1)[d]}

\def\xscale{1}
\def\yscale{1}

\def\Ellipse#1(#2)[#3,#4]{\def\xscale{#3}\def\yscale{#4}\relax
\Circle#1(#2)\def\xscale{1}\def\yscale{1}}

\def\dashEllipse(#1)[#2,#3]{\def\xscale{#2}\def\yscale{#3}\relax
\dashcirc(#1)\def\xscale{1}\def\yscale{1}}

\def\waveEllipse(#1)[#2,#3]{\def\xscale{#2}\def\yscale{#3}\relax
\wavecirc(#1)\def\xscale{1}\def\yscale{1}}

\def\halfEllipse#1(#2)[#3][#4,#5]{\def\xscale{#4}\def\yscale{#5}\relax
\halfcirc#1(#2)[#3]\def\xscale{1}\def\yscale{1}}

\def\halfdashEllipse(#1)[#2][#3,#4]{\def\xscale{#3}\def\yscale{#4}\relax
\halfdashcirc(#1)[#2]\def\xscale{1}\def\yscale{1}}

\def\halfwaveEllipse(#1)[#2][#3,#4]{\def\xscale{#3}\def\yscale{#4}\relax
\halfwavecirc(#1)[#2]\def\xscale{1}\def\yscale{1}}

\catcode`\@=\the\CatcodeOfAtSign

\begin{flushright}
Preprint SSU-HEP-98/04\\
Samara State University
\end{flushright}

\vspace{30mm}

\begin{center}
{\Large \bf Evaluation of hadronic vacuum polarization\\ contribution
to muonium hyperfine splitting}\\

\vspace{4mm}

R.N.~Faustov \\Scientific Council for Cybernetics,\\
Russian Academy of Sciences\\
117333, Moscow, Vavilov, 40, Russia,\\
A.~Karimkhodzhaev\\Department of Physics, Tashkent State University,\\
700095, Tashkent, Uzbekistan,\\
A.P.~Martynenko\\ Department of Theoretical Physics, Samara State University,\\
443011, Samara, Pavlov, 1, Russia
\end{center}

\begin{abstract}
The contribution of hadronic vacuum polarization (HVP) to the hyperfine splitting
of the muonium ground state is evaluated with the account of modern
experimental data on the cross section of $e^+e^-\rightarrow$ annihilation
into hadrons.

\end{abstract}

\newpage

In the framework of the quasipotential method the
particle interaction operator for $1\gamma$- and $2\gamma$-
processes takes the form \cite{DTF,STY,FK}:
\begin{equation}
V=V_{1\gamma}+V_{2\gamma}=V^c+\Delta V,
\end{equation}
\begin{displaymath}
V_{1\gamma}=T_{1\gamma},~~~V_{2\gamma}=T_{2\gamma}-T_{1\gamma}\times G^f\times T_{1\gamma},
\end{displaymath}
where $V^c$ is Coulomb potential, $T_{1\gamma}$, $T_{2\gamma}$ are one- and two-photon
scattering amplitudes off the energy shell,
$[G^f]^{-1}=(b^2-\vec p^2)/2\mu_R$ is the free two-particle Green function
($\mu_R$ is relativistic reduced mass) \cite{MF}.
Then the corresponding correction in the energy spectrum may be written as
follows:
\begin{equation}
\Delta E=<\psi^c_n|\Delta V|\psi^c_n>,
\end{equation}
where $\psi^c_n$ is the ordinary Coulomb wave function. The hadronic vacuum
polarization contribution to the energy spectrum is determined by the Feynman
diagrams represented on figure by means (1). To take into account vacuum polarization
in these one-loop diagrams, we must do the following substitution in the photon
propagator \cite{STY}
\begin{equation}
\frac{1}{k^2+i\varepsilon}\rightarrow\left(\frac{\alpha}{\pi}\right)^2\int_
{s_{th}}^\infty\frac{\rho(s)ds}{k^2-s+i\varepsilon},
\end{equation}
where the spectral function $\rho(s)$ is connected with known cross section of
$e^+e^-$ - annihilation into hadrons $\sigma^h$:
\begin{equation}
\rho(s)=\frac{R(s)}{3s}=\frac{\sigma^h(e^+e^-\rightarrow hadrons)}{3s\sigma_{\mu\mu}
(e^+e^-\rightarrow\mu^+\mu^-)},
\end{equation}
and $\sigma_{\mu\mu}(e^+e^-\rightarrow\mu^+\mu^-)=4\pi\alpha^2/3s$ is the
$e^+e^-$- annihilation cross section to muonic pair. Introducing Feynman's
parameterization (parameter $\lambda$), we can write the contribution of
HVP to muonium ground state hyperfine splitting in the form \cite{STY}:
\begin{equation}
\Delta E^h_{hfs}(\mu e)=\left(\frac{\alpha}{\pi}\right)^2m_em_\mu\tilde E_F
\int_{4m_\pi^2}^\infty\frac{R(s)ds}{3s}\int_0^1\frac{16-6\lambda-\lambda^2}
{m^2_\mu\lambda^2+s(1-\lambda)}d\lambda,
\end{equation}
$m_e, m_\mu, m_\pi$ are the masses of electron, muon and charged pion,
$\tilde E^F=8\alpha^4m^2_em^2_\mu/3(m_e+m_\mu)^3$ is the Fermi energy of
the muonium. The inner integral in (5) can be calculated analytically. For
the calculation of external integral we have used the parameterization of
R(s), which accounts the modern experimental data on annihilation cross section
of $e^+e^-$ into hadrons \cite{D,B,A,E,Bl,Ak,EJ,MD,Sw}.
\begin{figure}
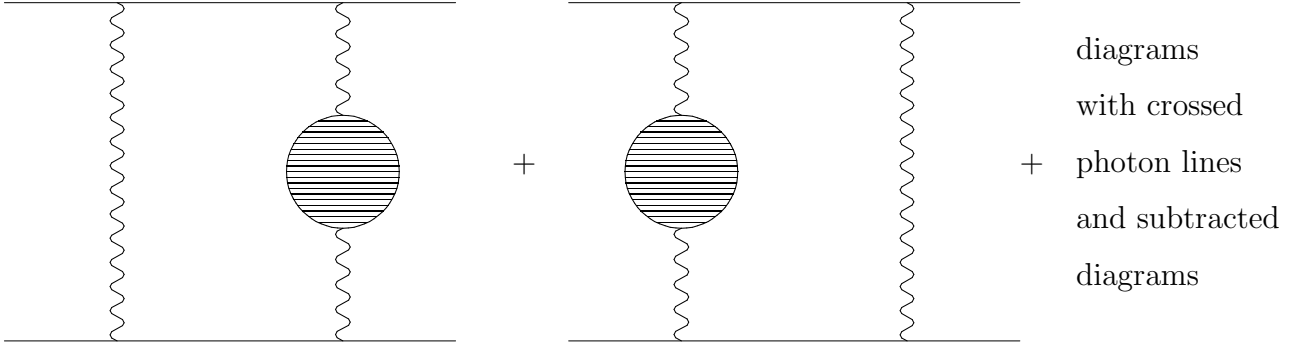

\magnitude=2000
\GRAPH(hsize=15){
\mov(0,0){\lin(4,0)}%
\mov(0,3){\lin(4,0)}%
\mov(5,0){\lin(4,0)}%
\mov(5,3){\lin(4,0)}%
\mov(3.,1.5){\Circle*(1)}%
\mov(6.,1.5){\Circle*(1)}%
\mov(4.5,1.5){+}%
\mov(9.,1.5){+}%
\mov(9.5,0.5){diagrams}%
\mov(9.5,2.){with crossed}%
\mov(9.5,1.5){photon lines}%
\mov(9.5,1.){and subtracted}%
\mov(9.5,2.5){diagrams}%
\mov(1,0){\wavelin(0,3)}%
\mov(3,0){\wavelin(0,1)}%
\mov(3,2){\wavelin(0,1)}%
\mov(6,0){\wavelin(0,1)}%
\mov(6,2){\wavelin(0,1)}%
\mov(8,0){\wavelin(0,3)}%
}

\vspace{10mm}

\caption{Feynman diagrams, defining the hadronic vacuum polarization
contribution to hyperfine structure of atom ($\mu e$)}
\end{figure}

The cross section $\sigma^h$ may be expressed as a sum of terms \cite{STY,FK}:
\begin{equation}
\sigma^h=\sigma_{\pi\pi}^h+\sigma^h_{res}+\sigma^h_{background}.
\end{equation}

The main
contribution to the cross section $\sigma^h$ is determined by the process
$e^++e^-\rightarrow \pi^++\pi^-$. The cross section of this reaction is
proportional to the modulus squared of pion form factor $F_\pi$.
The corresponding spectral function (4) may be presented
in the form:
\begin{equation}
\rho_{\pi\pi}(s)=\frac{(s-4m_\pi^2)^{3/2}}{12s^{5/2}}|F_\pi(s)|^2,
\end{equation}
Contrary to our recent calculation \cite{MF1}
to perform integrations in (5) for this case we have used the expression of
pion form factor, obtained in \cite{GP}:
\begin{equation}
F_\pi(s)=\frac{m_\rho}{m^2_\rho-s-im_\rho\Gamma_\rho(s)}\exp\left\{-\frac{s}{96\pi^2f^2_\pi}
Re\left[A\left(\frac{m^2_\pi}
{s},\frac{m^2_\pi}{m^2_\rho}\right)+\frac{1}{2} A\left(\frac{m_K^2}{s},
\frac{m_K^2}{m^2_\rho}\right)\right]\right\}
\end{equation}
where the function
\begin{equation}
A\left(\frac{m_P^2}{s},\frac{m_P^2}{\mu^2}\right)=\ln\left(\frac{m_P^2}{\mu^2}\right)+
\frac{8m_P^2}{s}-\frac{5}{3}+\sigma^3_P\ln\left(\frac{\sigma_P+1}{\sigma_P-1}\right),~~
\sigma_P=\sqrt{1-4m_P^2/s},
\end{equation}
\begin{equation}
\Gamma_\rho(s)=-\frac{m_\rho s}{96\pi^2 f_\pi^2} Im\left[
A\left(\frac{m^2_\pi}
{s},\frac{m^2_\pi}{m^2_\rho}\right)+\frac{1}{2} A\left(\frac{m_K^2}{s},\frac{m_K^2}
{m^2_\rho}\right)\right],
\end{equation}
where $f_\pi=92.42\pm 0.07\pm 0.25$ MeV \cite{PDG}, $m_K$, $m_\rho$ are
the masses of $K^{\pm}$ and $\rho$ mesons. This pion form factor gives good
description of the experimental data up to the energies of the order of 1 GeV
in terms of $m_\pi$, $m_K$, $m_\rho$ and $f_\pi$. To obtain the resonance
contribution to $\sigma^h$ the Breit-Wigner representation has been exploited.
The contribution of $\sigma^h_{background}$ was found by fitting of the
experimental points \cite{D,B,A,E,Bl,Ak,EJ,MD,Sw}. For description of the
experimental data we have used polinomial fit: $R(s)=\sum_{k=0}^n c_k(\sqrt{s})^k $
of high order on n. The coefficients $c_k$ were determined by means of the
least-squares method. The region of $\sqrt{s}>60$ GeV was described by QCD
formula, following from the assumption of the asymptotic freedom with six
quarks. The results of our calculations are presented in the table. The total
contribution of HVP to hyperfine muonium splitting is equal to:
\begin{equation}
\Delta E^h_{hfs}(\mu e)=(3.5988\pm 0.1045)\left(\frac{\alpha}{\pi}\right)^2
\frac{m_em_\mu}{m^2_\pi}\tilde E^F=0.2397\pm 0.0070~~kHz.
\end{equation}
This value is in accord with the result obtained in \cite{FK}: $\Delta E^h_{hfs}=
0.250\pm 0.016$ kHz. The more than twice higher precision of (11) is
explained by the account of new experimental data and by more accurate data
treatment \cite{EJ}. The new value (11) allows to increase the accuracy of
testing
radiative corrections to the ground state hyperfine splitting  in muonium.
It is important because a precision of
hyperfine muonium splitting measurement in the new Los Alamos experiment is
of order of $10^{-2}$ kHz.
Much of theoretical efforts was spent in the last
years in order to calculate new contributions to the muonium hyperfine
splitting of a high degree in $\alpha$ and $m_e/m_\mu$ \cite{KN,KN1,EGS}.
Theoretical value $\Delta E^{th}_{hfs}=4463302.69$ (1.34) (0.21) (0.16) kHz
\cite{KN}, taking into account (11), is in good agreement with
experimental result $\Delta E^{exp}_{hfs}=4463302.88$ (16) kHz \cite{DTF}.
The first and second errors in $\Delta E^{th}_{hfs}$ reflect the uncertainties
in the measurements of $\alpha^{-1}$ \cite{TK} and $m_\mu/m_e$. The third
error is purely theoretical and dominated by the coefficients in the
$\alpha (Z\alpha)^2$, $\alpha (Z\alpha)^3 \ln (Z\alpha)$ corrections \cite{KN1}.

\vspace{5mm}

\begin{tabular}{|c|c|c|}  \hline
final state &energy range $\sqrt{s}$, GeV & $\Delta E_{hf.s}^h(\mu e)/
\left(\frac{\alpha}{\pi}\right)^2\frac{m_em_\mu}{m_\pi^2}\tilde E^F$\\ \hline
$\rho,\omega\rightarrow 2\pi$ &(0.28, 0.81)&$ 2.0518\pm0.0134$ \\
$\omega\rightarrow 3\pi$ &(0.42, 0.81)& $0.1869\pm 0.0061$ \\
$\phi  $ & &$0.2122\pm 0.0077$	\\
$J/\Psi$ & & $0.0538\pm 0.0059$  \\
$\Upsilon$  & & $0.0006\pm 0.0001$ \\
hadrons &(0.81, 1.4)   &$0.5562\pm 0.0260$  \\
hadrons &(1.4, 2.2)   &$0.2406\pm 0.0199$  \\
hadrons &(2.2, 3.1)   &$0.1252\pm 0.0110$ \\
hadrons &(3.1, 5.0)   &$0.0906\pm 0.0122$  \\
hadrons &(5.0, 10.0)  &$0.0560\pm 0.0015$   \\
hadrons &(10.0, 40.0)&$0.0227\pm 0.0007$    \\
hadrons &(40.0, 60.0)&$0.0012\pm 0.0001$   \\
hadrons &$\sqrt{s}\geq 60.0$ &	0.0010	\\  \hline
   &Summary contribution   & $3.5988\pm 0.1045$   \\   \hline
\end{tabular}

\vspace{5mm}

We are grateful to Prof. T.~Kinoshita for suggestion to update the
calculation of HVP contribution to muonium hyperfine splitting
and to Prof. F.~Jegerlehner for helpful discussions of experimental data.
This work was supported by the Russian Foundation for Basic Research
(grant no. 98-02-16185) and the Program "Universities of Russia" (grant 2759).

\end{document}